\documentclass[a4paper]{article}

\usepackage[utf8]{inputenc}

\usepackage{graphicx}
\usepackage{hyperref}
\hypersetup{
	setpagesize  = false,
	colorlinks   = true,    
	urlcolor     = blue,    
	linkcolor    = black,   
	citecolor    = red    
}

\usepackage{authblk}
\usepackage{amsmath}
\usepackage{amssymb}
\usepackage{float}
\DeclareMathOperator{\sign}{sign}

\usepackage{xcolor}

\newcommand{\keywordsenglishname}{Keywords}

\renewenvironment{abstract}{%
	\begin{center}
		\begin{minipage}{14cm}
			{\textbf{\abstractname:}}
		}{
		\end{minipage}
	\end{center}
}

\title {Electric Motor -- SimuFísica\textsuperscript{\textregistered}: an application for teaching electromagnetism}
\author{Marco P. M. de Souza\thanks{Correspondence address: marcopolo@unir.br} $^1$ }
\author{Sidnei P. Oliveira$^{1,2}$}
\author{Valdenice L. Luiz$^{1}$}
\affil{{\small $^1$Departamento de Física, Universidade Federal de Rondônia, Ji-Paraná, Rondônia, Brazil.}}
\affil{{\small $^2$Secretaria de Estado da Educação -- SEDUC, Rondônia, Brazil.}}
\date{}

\usepackage{fancyhdr}
\begin{document}
	
	\maketitle
	\vspace{6pt}
	
\begin{abstract}
	In this work, we present the Electric Motor simulator, an application from the SimuFísica\textsuperscript{\textregistered} platform designed for classroom use. We briefly describe the technologies behind the application, the equations that govern its operation, some studies showing the dynamics of the electric motor, and, finally, the use of the application in High School and Higher Education.
	
	\vspace{0.5cm}
	\textbf{\textit{keywords}}: Electric motor, Computational simulation, Application, Simulator, Electromagnetism.
\end{abstract}

	\section{Introduction}
	
	The topic of the electric motor can be an effective starting point for teaching electromagnetism due to the various topics related to its operation. Since the electric motor involves the principles of electromagnetism, students can directly see how the laws of this theory are applied to a device present in our daily lives, making the subject more relevant. The study of the electric motor requires the understanding and integration of several electromagnetism concepts, such as electric current, magnetic field, magnetic flux, magnetic force, and Faraday's law, as well as mechanics concepts, such as torque, moment of inertia, and angular acceleration.
	
	Experiments involving the application of the electric motor in physics teaching have been a recurring theme in the literature, appearing in various articles \cite{Yap, Schubert, Hudha, Diniz, Pires, Monteiro} and dissertations \cite{Silva, Euzebio}. There are also studies related to the numerical solution of the dynamics of the electric motor \cite{Chiasson, Moreno}. On the other hand, computational simulations of an electric motor for educational purposes are quite scarce. This motivated us to develop and present the Electric Motor application, available on the SimuFísica\footnote{\url{https://simufisica.com/en/}} platform, aimed at teaching high school and higher education, which we will address in the following sections.
	
	The Electric Motor application presented here offers the user a simulation, significantly distinguishing itself from videos and animations. As a learning object \cite{Arantes}, its main advantage over internet videos about electric motors is the presence of interactivity, where the user can configure parameters and initial conditions, as addressed in Section \ref{app-motor-eletrico}. Regarding the animations found on the web \cite{oPhysics, JavaLab}, even those that claim to be simulations and offer a certain degree of user interaction are quite limited, as they are unlikely to represent all the physical phenomena that the real-time solution of differential equations can illustrate, such as the effect of Faraday's law on the coil current, for example.

	\section{The SimuFísica\textsuperscript{\textregistered} Platform}
	
	SimuFísica\textsuperscript{\textregistered} is a collection of simulator applications aimed at teaching Physics at high school and higher education levels. With its multilingual and multiplatform nature, SimuFísica\textsuperscript{\textregistered} offers simulations covering topics such as electric energy consumption, the ideal gas, and the propagation of the wave function in quantum mechanics, for example. These simulators can be accessed online or installed on computers, tablets, and smartphones running various operating systems. Access to the SimuFísica platform, which is free and ad-free software, can be done both through the web at \url{https://simufisica.com/en/} and by downloading it from the Google Play, App Store, Microsoft Store, and Snap Store.
	
	Aimed at students and teachers, the use of SimuFísica\textsuperscript{\textregistered} in the classroom, combined with good planning, can bring significant advantages to Physics teaching \cite{Cristiane}. The platform offers the opportunity to visualize physical phenomena interactively, making the understanding of abstract concepts easier. Additionally, students can conduct virtual experiments, exploring different parameters and quickly observing the consequences. This promotes active learning, stimulates critical thinking, and develops problem-solving skills. The immediate feedback provided by SimuFísica\textsuperscript{\textregistered} enables students to correct mistakes, enhancing their learning process. The platform also offers flexible access, being available online and as an app for mobile and desktop devices, allowing students and teachers to use the simulators anywhere and at any time. By incorporating SimuFísica\textsuperscript{\textregistered} into lesson plans, teachers can enrich classroom instruction, complement theory with practical examples, and help students visualize real-world applications of Physics.

	\section{The Electric Motor Application}
	\label{app-motor-eletrico}
	
\begin{figure}[ht]
		\centering
		\includegraphics[width=0.7\linewidth]{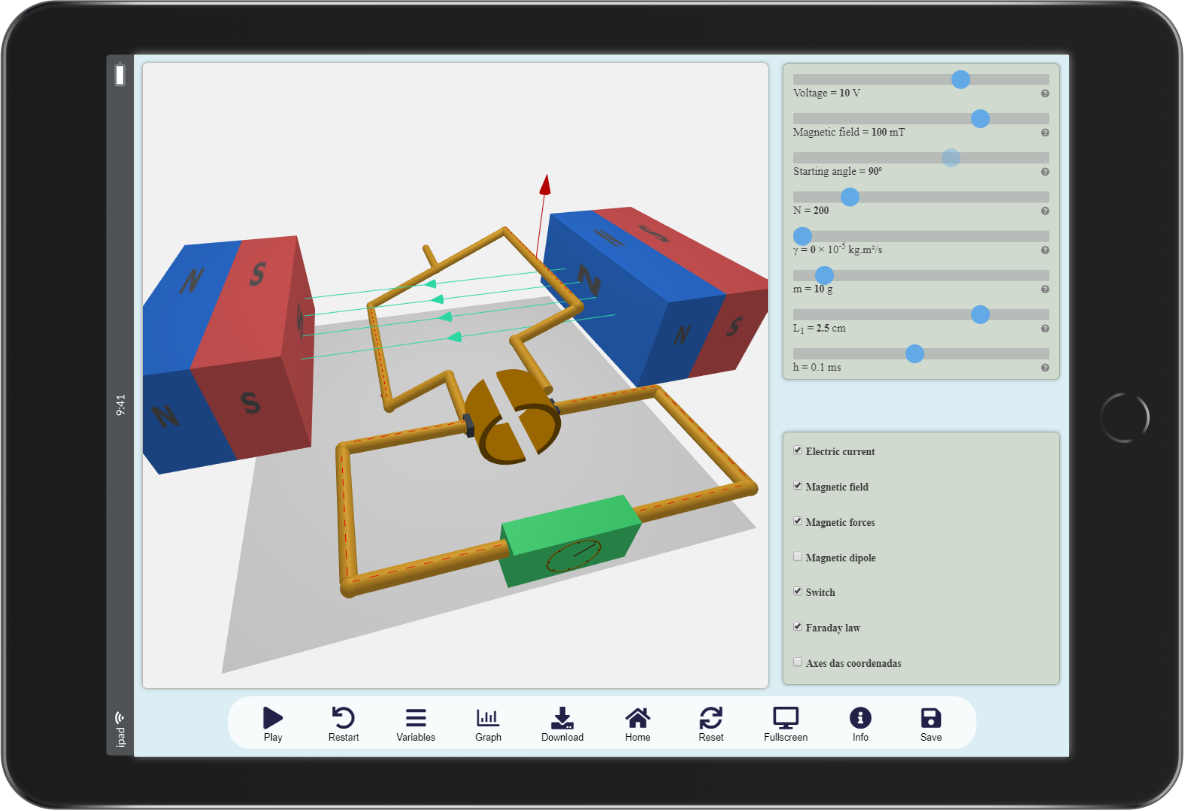}
		\caption{Version 1.9 of the Electric Motor application running on an iPad 6. The red and green arrows represent, respectively, the magnetic forces acting on each loop and the magnetic field lines between the magnets. Available through the link \url{https://simufisica.com/en/electric-motor/}.}
		\label{fig1}
	\end{figure}

	This is a simulator of a direct current electric motor composed of a coil with rectangular loops in the presence of a uniform magnetic field generated by a pair of magnets (Fig. \ref{fig1}). The application is highly interactive, allowing the user to configure various variables. The user can adjust the voltage of the power source supplying the coil, the intensity and direction of the magnetic field, the initial angle between the coil plane and the magnetic field, the number of loops, the damping constant acting on the coil's rotation, the coil mass, the loop width, and the numerical integration step for the system of ordinary differential equations (ODEs) governing the system's dynamics (see Sec. \ref{simulacao-motor-eletrico}).
	
	In the simulator, the user has the option to view some variables numerically, such as the coil's rotational frequency and the magnetic flux through the loops, as well as their time evolution graphs. The application also provides a representation of the electric current, the magnetic force vector acting on the loops, and the magnetic field lines passing through them.
	
	The decision to develop an application for simulating an electric motor was influenced by several factors. In addition to the motivations presented in the introduction, we have the fact that understanding the operation of electric motors is one of the skills required by Specific Competency 1 of the Brazilian National Common Curriculum Base (BNCC) in the area of Natural Sciences and its Technologies for High School \cite{BNCC}. We will revisit this point in Section \ref{ensino-fisica}.
	
	Regarding the two types of electric motors based on the type of power source—direct current or alternating current—we opted to simulate the former due to its simplicity. These motors often operate at low voltages, such as 3 V, and can be quite small, making them easier to use in the classroom. Additionally, the construction of this type of motor is relatively simple, allowing high school and undergraduate students to easily replicate it \cite{Silva}. Lastly, it is worth noting that direct current topics are usually introduced to students well before alternating current, which allows for the earlier introduction of the direct current electric motor in the classroom.
	
	The online version of the Electric Motor app, like all others on the SimuFísica\textsuperscript{\textregistered} platform, has its source code based on three major web technologies: HTML5 (content, i.e., HTML tags), CSS (layout), and JavaScript (dynamics, essentially). JavaScript is responsible for solving the ODEs based on the parameters and initial conditions defined by the user, rendering the simulation in real time on the screen through manipulation of the Canvas tag, and adapting the application to devices of various sizes, such as desktops, tablets, and smartphones. The 3D drawing of the electric motor is carried out through the Three.js library \cite{three-js}.
	
\begin{figure}[ht]
		\centering
		\includegraphics[width=0.6\linewidth]{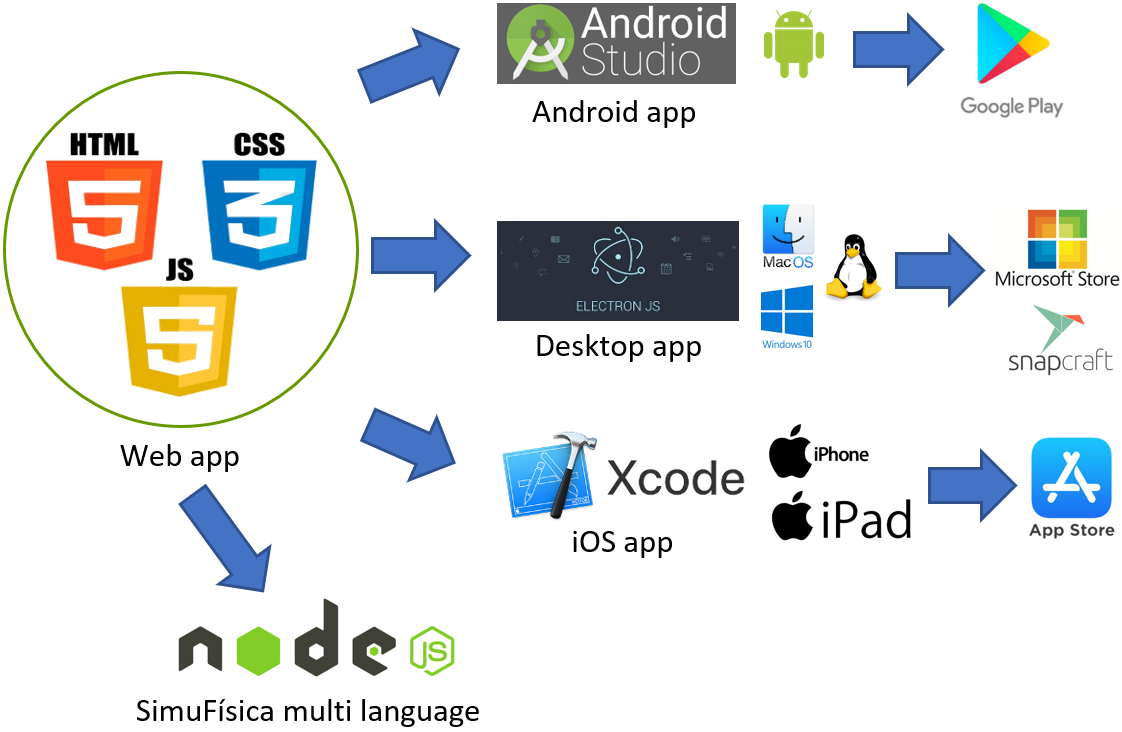}
		\caption{Simplified development flowchart of the app.}
		\label{fig2}
	\end{figure}

	The English and Spanish versions are obtained through translation, which is primarily done automatically and locally (on the development computer) using the Node.js framework and the Google Translate API (application programming interface), available for free via the npm (Node Package Manager) package manager. The downloadable versions, available on the platform or through app stores, are developed using the Android Studio IDE (Google Play Store) and Xcode (Apple App Store), or with the Electron.js framework (Microsoft Store $\rightarrow$ Windows, and Snap Store $\rightarrow$ Linux). Fig. \ref{fig2} shows a simplified illustration of the main technologies involved in developing the various versions.
	
	\section{The Simulation of the Electric Motor}
	\label{simulacao-motor-eletrico}
	
\begin{figure}[ht]
		\centering
		\includegraphics[width=0.35\linewidth]{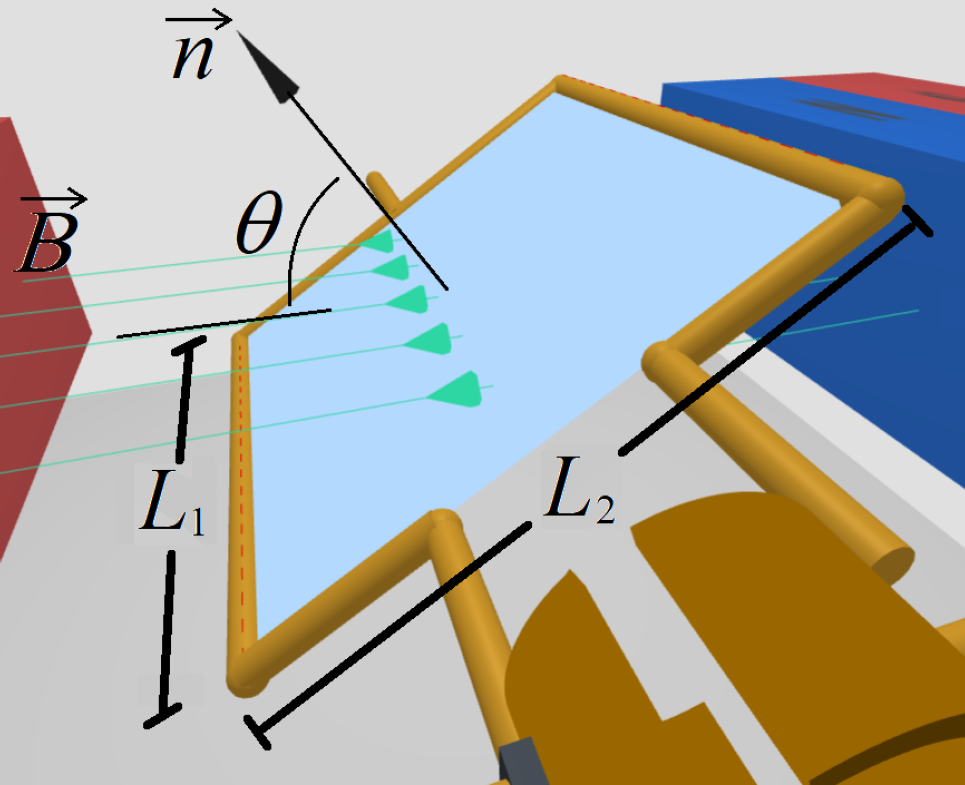}
		\caption{Some variables and parameters involved in the electric motor simulation. $\theta$ is the angle between the normal vector $\vec{n}$ to the plane of the coil and the magnetic field vector $\vec{B}$.}
		\label{fig3}
	\end{figure}

	The electric motor idealized in the application consists of a coil of mass $m$ with $N$ rectangular turns of dimensions $L_1\times L_2$, where $L_2 = 2.5$ cm is fixed, while $L_1$, referring to the side always perpendicular to the magnetic field, can be adjusted by the user (see Fig. \ref{fig3}). The magnetic field $\vec{B}$ is considered uniform throughout the region occupied by the coil. We consider that the complete electric motor system has a total electrical resistance of $R = 10$ $\Omega$.
	
	What makes the coil start to rotate is the action of magnetic forces on the sides of dimension $L_1$ of the turns, which generate a torque on the coil, since in the absence of torque (net zero torque), the coil keeps rotating, as the effect of the torque is to change the angular momentum. The magnitude of each of the total magnetic forces acting on the sides of dimension $L_1$, represented by the red arrows in the application (Fig. \ref{fig1}), is given by \cite{Halliday}
	
	\begin{equation}
		\label{forca-magnetica}
		F_B = i N L_1 B,
	\end{equation}
	
	\noindent where $i$ is the current in the coil.
	
	As we are dealing with a problem related to the rotation of a body, we apply Newton's second law for rotations. Starting from Eq. (\ref{forca-magnetica}), it is possible to deduce that the magnitude of the torque responsible for the rotation of the coil is given by \cite{Halliday}
	
	\begin{equation}
		\label{torque-bobina}
		\tau = i N L_1 L_2 B \sin{\theta},
	\end{equation}
	
	\noindent where $\theta$ is the angle between the normal vector to the coil plane and the magnetic field.
	
	The moment of inertia $I$ of each turn relative to the axis of rotation can be calculated from its definition in mechanics \cite{Moyses1}:
	
	\begin{equation}
		\label{momento-inercia}
		I = \int \rho^2 dm,
	\end{equation}
	
	\noindent where $\rho$ is the distance of each mass element $dm$ of the body to the axis of rotation. The combined moment of inertia of the two sides of dimension $L_1$, each with mass $m_1$, is given by
	
	\begin{equation}
		\label{momento-inercia-L1}
		I_1 = 2\times\dfrac{m_1L_2^2}{4},
	\end{equation}
	
	\noindent since all points on the $L_1$ side are at the same distance $\rho = L_2/2$ from the axis of rotation. The combined moment of inertia of the two sides of dimension $L_2$, each with mass $m_2$, calculated from Eq. (\ref{momento-inercia}), is given by \cite{Moyses1}
	
	\begin{equation}
		\label{momento-inercia-L2}
		I_2 = 2\times\dfrac{m_2L_2^2}{12}.
	\end{equation}
	
	\noindent Using Eqs. (\ref{momento-inercia-L1}) and (\ref{momento-inercia-L2}), we obtain the moment of inertia of the coil:
	
	\begin{equation}
		\label{momento-inercia-espira}
		I_M = I_1 + I_2 = \dfrac{mL_2^2}{L_1+L_2} \left( \dfrac{L_1}{4} + \dfrac{L_2}{12}  \right),
	\end{equation}
	
	\noindent where we consider that the coil has a uniform mass distribution and that $m = (m_1 + m_2)$.
	
	The current $i$ that appears in the torque [Eq. (\ref{torque-bobina})] is not constant, since with the beginning of the coil's rotation, we have two additional voltages arising from Faraday's law, which act together with the external power supply voltage $V_0$: the voltage due to the variation of the magnetic flux ($\mathcal{E}_1$) and the voltage from the variation of the current in the coil ($\mathcal{E}_2$). From Faraday's law, we write $\mathcal{E}_1$ as \cite{Moyses3}
	
	\begin{equation}
		\label{lei-de-faraday}
		\mathcal{E}_1 = -\dfrac{d\Phi_B}{dt},
	\end{equation}
	
	\noindent where $\Phi_B$ is the magnetic flux through the coil. Since we are considering the magnetic field uniform inside the coil, we write $\Phi_B$ as \cite{Moyses3}
	
	\begin{equation}
		\label{fluxo-magnetico}
		\Phi_B = -NL_1L_2B\cos{\theta}.
	\end{equation}
	
	\noindent Thus, we rewrite the induced electromotive force [Eq. (\ref{lei-de-faraday})] as
	
	\begin{equation}
		\label{lei-de-faraday2}
		\mathcal{E}_1 = -NL_1L_2B\omega\sin{\theta},
	\end{equation}
	
	\noindent where
	
	\begin{equation}
		\label{omega}
		\omega = \dfrac{d\theta}{dt}
	\end{equation}
	
	\noindent is the angular frequency of the coil's rotation.
	
	Regarding the effect of the induced electromotive force due to the current variation, we know that
	
	\begin{equation}
		\label{auto-inducao}
		\mathcal{E}_2 = -N^2L\dfrac{di}{dt},
	\end{equation}
	
	\noindent where $L$ is the self-inductance of each coil turn (the total inductance is proportional to the square of the number of turns \cite{Moyses3}). The value of the self-inductance for rectangular turns of dimensions $L_1 \times L_2$ composed of wires with a circular cross-section of radius $a$ can be calculated from the following equation:
	
	\begin{eqnarray}
		L &=& \dfrac{\mu_0}{\pi} \left[  L_1\ln\left( \dfrac{2L_1}{a} \right) + L_2\ln\left( \dfrac{2L_2}{a} \right) - \left( L_1 + L_2\right) \left( 2 - \dfrac{Y}{2}\right) +  \right.  \nonumber \\
		&+& \left.   2\sqrt{L_1^2 + L_2^2} - L_1\arcsin\left( \dfrac{L_1}{L_2} \right) - L_2\arcsin\left( \dfrac{L_2}{L_1} \right) + a    \right], 
	\end{eqnarray}
	
	\noindent where $\mu_0$ is the vacuum magnetic permeability, and $Y$ is a parameter that depends on how the current flows through the wire ($Y = 0$ if the current flows on the wire surface and $Y = 1/2$ if it flows uniformly throughout its cross-section) \cite{Rosa, Dengler}.
	
	Using Kirchhoff's loop law \cite{Halliday} in the circuit composed of the power supply and the coil, and using Eqs. (\ref{lei-de-faraday2}) and (\ref{auto-inducao}), we arrive at the following ODE:
	
	\begin{equation}
		\label{lei-das-malhas}
		N^2L\dfrac{di}{dt} = V_0\sign(\sin\theta) - Ri - NL_1L_2B\omega\sin{\theta}.
	\end{equation}
	
	\noindent The term $\sign(\sin\theta)$, where $\sign(x)$ is the sign of any number $x$, appears in Eq. (\ref{lei-das-malhas}) to account for the effect of the commutator on the dynamics of the electric motor \cite{Moreno}, a device that reverses the direction of current in the coil, making the DC motor work. That is, whenever $0 \leqslant \theta < \pi$, we have a voltage $V_0$ from the power supply acting on the coil, while if $\pi \leqslant \theta < 2\pi$, this voltage is $-V_0$.

	The relationship between the angular acceleration of the coil and the resulting torque exerted on it is given by Newton's second law for rotations:
	
	\begin{equation}
		\label{2a-lei-newton}
		I_M\dfrac{d\omega}{dt} = \tau - \gamma\omega,
	\end{equation}
	
	\noindent where $\gamma$ is a damping constant that leads to a dissipative torque proportional to the angular velocity of the coil. This term was included to account for the kinetic energy loss of the motor due to friction present in the brushes and bearings, for example \cite{Chiasson}.
	
	By combining equations (\ref{torque-bobina}), (\ref{omega}), (\ref{lei-das-malhas}), and (\ref{2a-lei-newton}), we arrive at the following system of ODEs \cite{Chiasson}:
	
	\begin{subequations}
		\label{edos1}
		\begin{align}
			\dfrac{d\theta}{dt} &= \omega\\
			N^2L\dfrac{di}{dt} &= V_0\sign(\sin\theta) - Ri - NL_1L_2B\omega\sin{\theta}\\
			I_M\dfrac{d\omega}{dt} &= i N L_1 L_2 B \sin{\theta} - \gamma\omega.
		\end{align}
	\end{subequations}
	
	In the app Electric Motor from the SimuFísica platform, we consider, as an approximation, that $N^2Ldi/dt = 0$. In other words, we are neglecting the effect of the induced voltage in the coil due to the variation of the current in it. Note that for a square loop of dimensions $2.5 \times 2.5$ cm$^2$, we have $L \sim 10^{-8}$ H for a wire with a diameter of 2 mm \cite{Dengler}, which implies that the term $N^2Ldi/dt$ is only relevant during the short time interval of the commutator’s action, which can lead to a sudden variation in the current in the coil. Thus, setting $N^2Ldi/dt = 0$ in Eq. (\ref{edos1}b), we obtain an expression for the value of the current in terms of the other variables:
	
	\begin{equation}
		\label{corrente}
		i = \dfrac{V_0\sign(\sin\theta)}{R} - \dfrac{NL_1L_2B}{R}\omega\sin{\theta}
	\end{equation}
	
	Inserting Eq. (\ref{corrente}) into Eq. (\ref{edos1}c), we arrive at the following system of ODEs:
	
	\begin{subequations}
		\label{edos2}
		\begin{align}
			\dfrac{d\theta}{dt} &= \omega\\
			\dfrac{d\omega}{dt} &= a_1|\sin{\theta}| - a_2\omega\sin^2{\theta} - a_3\omega,
		\end{align}
	\end{subequations}
	
	\noindent where $a_1 = NL_1L_2V_0B/(I_MR)$, $a_2 = (NL_1L_2B)^2/(I_MR)$, and $a_3 = \gamma/I_M$. In the application, this system of ODEs [equations (\ref{edos2})] is numerically integrated by the fourth-order Runge-Kutta method \cite{Press} with time step $h$, which can be chosen by the user. It is important to highlight that the fact that the system of ODEs (\ref{edos2}) is much simpler than the system (\ref{edos1}) makes it possible for the simulator to work well even on computers and other devices with more modest processing power.
	
\begin{figure}[ht]
		\centering
		\includegraphics[width=0.6\linewidth]{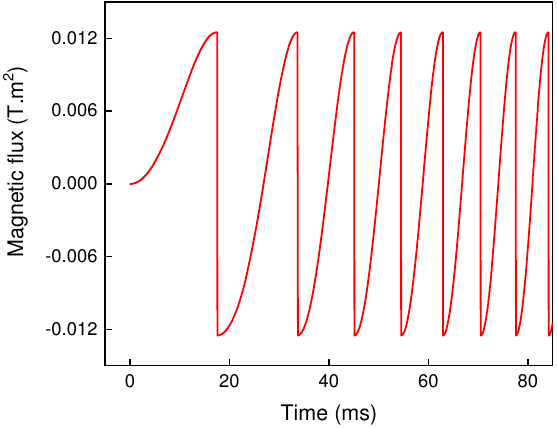}
		\caption{Magnetic flux in the loops.}
		\label{fig4}
	\end{figure}

	In Fig. \ref{fig4}, we have the time evolution of the total magnetic flux through the loops. We consider a coil with $N = 200$ turns, power supply voltage $V_0 = 10$ V, magnetic field $B = 100$ mT, initial angle relative to the magnetic field $\theta_0 = \pi/2$ rad, $L_1 = 2.5$ cm, total mass $m = 10$ g, and damping coefficient $\gamma = 0$. We use $h = 0.05$ ms as the Runge-Kutta integration step. From the figure, we can observe a maximum magnetic flux with a magnitude of 0.0125 T.m$^2$. Around $t = 17.5$ ms, and also at certain later moments, a sudden change in the magnetic flux is observed, which comes from the action of the commutator. This device allows for the reversal of the current direction in the loops and, consequently, the magnetic dipole vector to be constantly ``trying'' to align with the magnetic field of the magnets, an effect that generates the rotation of the motor.
	
\begin{figure}[H]
		\centering
		\includegraphics[width=0.6\linewidth]{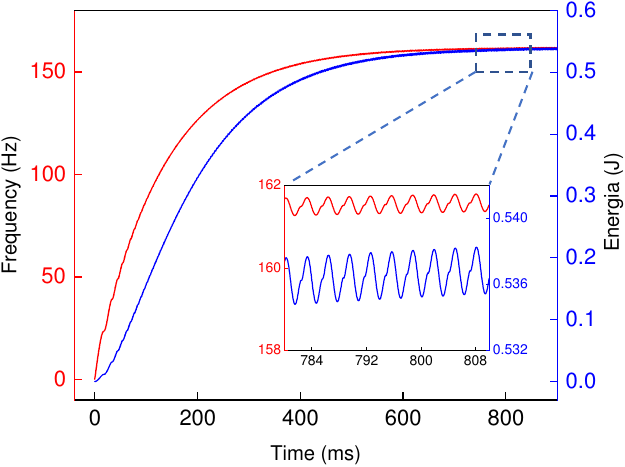}
		\caption{Rotational frequency and kinetic energy of the coil as a function of time. \textit{Inset}: zoom of the dashed rectangular region.}
		\label{fig5}
	\end{figure}

	We can also observe that the oscillation period of the magnetic flux presented in Fig. \ref{fig4} decreases over time. This depends on several parameters, such as the mass of the coil, the value of the damping coefficient, and the value of the magnetic field, which results in the increase in the average rotational frequency until it reaches the steady state (Fig. \ref{fig5}). In the case where $\gamma = 0$, the oscillation frequency does not increase indefinitely due to the opposing current induced in the coil by Faraday's law.
	
\begin{figure}[ht]
		\centering
		\includegraphics[width=0.5\linewidth]{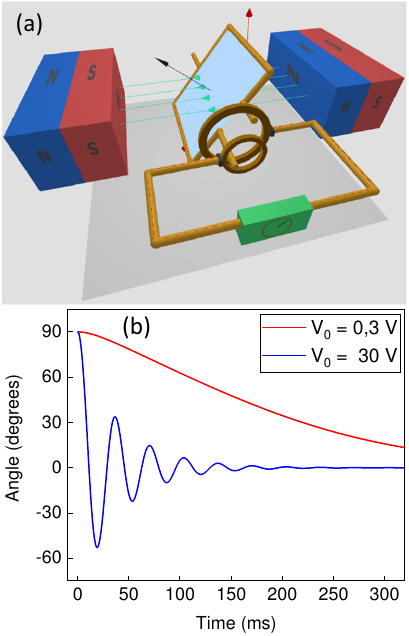}
		\caption{(a) Simulator with the commutator disabled. The black arrow represents the magnetic dipole vector of each loop. 
			(b) Coil angle $\theta$ as a function of time for two power supply voltages: 0.3 and 30 V.}
		\label{fig6}
	\end{figure}

	It is interesting to note the system's dynamics without the presence of the commutator, as in this case, it is possible to highlight the tendency of the magnetic dipoles to align with the magnetic fields. In the application, this can be done by disabling the \texttt{Switch On/Off} option. Setting $\gamma = 5\times 10^{-5}$ kg.m$^2$/s, keeping the other parameters and initial conditions indicated in the previous figures, and disabling the commutator in the application, we can observe in Fig. \ref{fig6} the torque experienced by the coil and its tendency to align with the magnetic field of the magnets. It is possible to observe, from Fig. \ref{fig6}(b), the presence of two well-known regimes of the simple harmonic oscillator \cite{Moyses2}: over-damped (red curve, where the power supply voltage is $V_0 = 0.3$ V) and under-damped (blue curve, where we have $V_0 = 30$ V).
	
\section{The Electric Motor App in Physics Education}
\label{ensino-fisica}

Since the app is compatible with mobile devices, it is possible to easily project the open simulation from the device to a projector without using video cables, by combining it with a TV box, such as Apple TV (iOS and iPadOS) or Roku Express (Android) \cite{Cristiane}, or through the projector's own wifi, if available. Below we present a commentary on the use of the Electric Motor app in both high school and higher education.

\subsubsection*{High School}

The use in high school classrooms can be anchored in the specific competencies of the field of Natural Sciences and its Technologies in the National Common Curricular Base (Base Nacional Curricular Comum -- BNCC -- Brazil) \cite{BNCC}. Among them, we can start with Specific Competence 1, related to natural phenomena and technological processes and their relationships with matter and energy. This includes topics from Physics, always in an interdisciplinary way, such as energy conservation, electric energy consumption, energy efficiency, electric motors, coils, batteries, and environmental impacts and sustainability \cite{BNCC}.

Specific Competence 3 is another directly linked to the use of the Electric Motor app in the classroom. This competence emphasizes the applications of scientific and technological knowledge and their impacts on contemporary society, including ethical, moral, cultural, social, political, and economic aspects. It highlights students' ability to interpret information in the form of text, equations, graphs, and tables, as well as to analyze the functioning of electrical and electronic equipment \cite{BNCC}.

All the competencies mentioned recommend the use of digital devices and apps \cite{BNCC} as supporting tools in teaching. A high school teacher can use the Electric Motor app either in isolation, emphasizing the Physics behind the motor's operation and the technological applications of electromagnetism, or in conjunction with other apps from the SimuFísica\textsuperscript{\textregistered} platform, such as ``Mechanical Energy Conservation,'' ``Electric Energy Consumption,'' and ``Mass Spectrometry,' for example.

\subsubsection*{Higher Education}

The physics of the electric motor is explored in the famous textbook ``Fundamentals of Physics - Volume 3 - Electromagnetism'' \cite{Halliday}, which is listed as essential or supplementary reading in several higher education courses, such as Physics, Mathematics, Chemistry, and various Engineering disciplines. Professors of these courses can consider the Specific Competencies mentioned in the previous subsection as motivation for a class involving topics related to the functioning of electric motors. Additionally, it is possible to focus on concepts and calculations involving magnetic force, the right-hand rule, magnetic flux, Faraday's law, and the magnetic dipole.

Besides the \texttt{Graph} button, which shows the evolution of some variables like angle, current, and magnetic flux, one can also use the information presented after clicking the \texttt{Variables} button. This allows the student or teacher to compare this information with results from calculations related to the magnetic force applied to a current-carrying wire and the magnetic flux through a coil, promoting learning.

\section{Final Considerations}

The Electric Motor app, part of the SimuFísica\textsuperscript{\textregistered} platform, can be a valuable tool in the teaching of electromagnetism at both high school and higher education levels. Due to its fast startup—always less than three seconds on all tested devices—and its compatibility with current smartphones, this simulator can be used very conveniently in academic environments, making lessons more engaging.

Several improvements are in development and will soon be implemented in the Electric Motor app. Among the main highlights are the option for the user to choose between a direct current or alternating current source to simulate the electric motor, as well as the functionality to operate the system in reverse, that is, as a current generator due to Faraday's law. The app will be used in high schools in Ji-Paraná and in the Physics courses at the Ji-Paraná campus of the Federal University of Rondônia (Brazil), so that bug fixes and improvements can be made after classroom use.

\section*{Acknowledgments}

Marco P. M. de Souza acknowledges the support from the National Council for Scientific and Technological Development -- CNPq (grant number 304017/2022-1). Sidnei P. Oliveira acknowledges the support from the Coordination for the Improvement of Higher Education Personnel -- CAPES -- Funding Code 001.

	\section*{A.I. disclosure statement}
	
	This article was translated from Portuguese using generative AI based on the version available at \url{https://doi.org/10.1590/1806-9126-RBEF-2023-0219}.


\begin{thebibliography}{99}
		\bibitem{Yap} J. Yap, D. MacIsaac, \textit{Phys. Educ.} \textbf{41}, 427 (2006).
		\bibitem{Schubert} T.F. Schubert, F.G. Jacobitz, E.M. Kim, \textit{IEEE Transactions on Education} \textbf{52}, 57 (2009).
		\bibitem{Hudha} M.N. Hudha \textit{et al}, \textit{J. Phys.: Conf. Ser.} \textbf{1318}, 012004 (2019).
		\bibitem{Diniz} A.M.F. Diniz, R.D. Araújo, \textit{Rev. Bras. Ensino Fís.} \textbf{41}, e20180216 (2019).
		\bibitem{Pires} C.F.J.S. Pires, P.C. Ferrari, J.R.O. Queiroz, \textit{R. Bras. de Ensino de C\&T} \textbf{6}, 29 (2013).
		\bibitem{Monteiro} I.C.C. Monteiro, M.A.A. Monteiro, J.S.E. Germano, A. Gaspar, \textit{Caderno Brasileiro de Ensino de Física} \textbf{27}, 371 (2010).
		\bibitem{Silva} J.A. Silva, \textit{O uso do motor elétrico para o ensino do eletromagnetismo}. Dissertação de Mestrado, Universidade Federal Rural de Pernambuco, Recife (2009).
		\bibitem{Euzebio} G.J. Euzébio, \textit{Motores elétricos como ideia âncora para a organização sequencial no ensino de eletricidade e magnetismo}. Dissertação de Mestrado, Universidade Federal Santa Catarina, Araranguá (2019).
		\bibitem{Chiasson} J. Chiasson, \textit{Modeling and High-Performance Control of Electric Machines} (\textit{John Wiley \& Sons, New York, 2005}).
		\bibitem{Moreno} E.F. Moreno, \textit{Eur. J. Phys.} \textbf{39}, 055203 (2018).
		\bibitem{Arantes} A.R. Arantes, M.S. Miranda, N. Studart, \textit{Física na Escola} \textbf{11}, 27 (2010).
		\bibitem{oPhysics}oPhysics: Interactive Physics Simulations. \url{https://ophysics.com/em10.html}.
		\bibitem{JavaLab}JavaLab: Science simulations. \url{https://javalab.org/en/dc_motor_en/}.
		\bibitem{Cristiane} C.M. Oliveira, \textit{Ensino da temática energia com três aplicativos da plataforma SimuFísica no Novo Ensino Médio}. Dissertação de Mestrado, Universidade Federal de Rondônia, Ji-Paraná, Brazil (2023).
		\bibitem{BNCC}MINISTÉRIO DA EDUCAÇÃO, \textit{Base Nacional Comum Curricular,  19 de dezembro de 2018,} \url{http://basenacionalcomum.mec.gov.br/images/BNCC_EI_EF_110518_versaofinal_site.pdf}.
		\bibitem{three-js} three.js. \url{https://threejs.org/}.
		\bibitem{Halliday} D. Halliday, R. Resnick e J. Walker, Fundamentos de Física -- Eletromagnetismo (\textit{LTC, Rio de Janeiro, 2010}), v. 3, 8 ed.
		\bibitem{Moyses1} H. Moysés Nussenzveig, \textit{Curso de Física Básica -- Mecânica} (Editora Blucher, São Paulo, 2002), v. 1, 4 ed.
		\bibitem{Moyses3} H. Moysés Nussenzveig, \textit{Curso de Física Básica -- Eletromagnetismo} (Editora Blucher, São Paulo, 2002), v. 3, 4 ed.
		\bibitem{Moyses2} H. Moysés Nussenzveig, \textit{Curso de Física Básica -- Fluidos, Oscilações e Ondas, e Calor} (Editora Blucher, São Paulo, 2002), v. 2, 4 ed.
		\bibitem{Rosa} E.B. Rosa, \textit{Bulletin of the Bureau of Standards} \textbf{4}, 301 (1907).
		\bibitem{Dengler} R. Dengler, \textit{Advanced Electromagnetics} \textbf{5}, 1 (2016).		
		\bibitem{Press} W.H. Press, Numerical Recipes in C: The Art of Scientific Computing (\textit{Cambridge University Press, New York, 2007}), 3 ed.
	\end{thebibliography}
\end{document}